\documentclass[plaindraft,letter]{jpp-AAS}

\usepackage{graphicx,amsmath}
\usepackage{natbib}
\usepackage[pdftex,colorlinks,citecolor=blue]{hyperref}
\usepackage[noabbrev]{cleveref}

\def\jcp{J.~Comput.~Phys.}

\def\apj{Astrophys.~J.}
\def\apjl{Astrophys.~J.~Lett.}
\def\aap{Astron.~Astrophys.}
\def\araa{Ann.~Rev.~Astron.~Astrophys.}
\def\mnras{Mon.~Not.~R.~Astron.~Soc.}
\def\nat{Nature}

\def\pop{Phys.~Plasmas}
\def\prl{Phys.~Rev.~Lett.}
\def\jgr{J.~Geophys.~Res.}
\def\jgrs{J.~Geophys.~Res. (Space Physics)}
\def\ssr{Space Science Rev.}

\def\grl{Geophys.~Res.~Lett.}
\def\pop{Phys.~Plasmas}
\def\pof{Phys.~Fluids}

\def\jpp{J.~Plasma Phys.}

\def\ropp{Rev.~Plasma Phys.}
\def\ppcf{Plasma Phys.~Control.~Fusion}

\ifCUPmtlplainloaded \else
  \checkfont{eurm10}
  \iffontfound
    \IfFileExists{upmath.sty}
      {\typeout{^^JFound AMS Euler Roman fonts on the system,
                   using the 'upmath' package.^^J}%
       \usepackage{upmath}}
      {\typeout{^^JFound AMS Euler Roman fonts on the system, but you
                   dont seem to have the}%
       \typeout{'upmath' package installed. JPP.cls can take advantage
                 of these fonts, if you use 'upmath' package.^^J}%
       \providecommand\upi{\pi}%
      }
  \else
    \providecommand\upi{\pi}%
  \fi
\fi

\ifCUPmtlplainloaded \else
  \checkfont{msam10}
  \iffontfound
    \IfFileExists{amssymb.sty}
      {\typeout{^^JFound AMS Symbol fonts on the system, using the
                'amssymb' package.^^J}%
       \usepackage{amssymb}%
       \let\le=\leqslant  
       \let\ge=\geqslant  
      }{}
  \fi
\fi

\ifCUPmtlplainloaded \else
  \IfFileExists{amsbsy.sty}
    {\typeout{^^JFound the 'amsbsy' package on the system, using it.^^J}%
     \usepackage{amsbsy}}
    {\providecommand\boldsymbol[1]{\mbox{\boldmath $##1$}}}
\fi

\DeclareMathAlphabet{\mathsfbi}{OT1}{\sfdefault}{bx}{sl}
\DeclareMathVersion{sfletters}
\SetSymbolFont{letters}{sfletters}{OML}{ntxsfmi}{b}{it}

\makeatletter
\newcommand{\mathbfsbilow}[1]{%
  \text{\mathversion{sfletters}$\m@th#1$}%
}

\DeclareRobustCommand{\tensor}[1]{%
  \begingroup
  \ifcat\noexpand #1\relax
    \edef\greek@test{\detokenize{#1}}%
    \edef\greek@test{\expandafter\@cdr\greek@test\@nil}%
    \edef\greek@test{\expandafter\@car\greek@test\@nil}%
    \edef\x{\the\lccode\expandafter`\greek@test}%
    \edef\y{\number\expandafter`\greek@test}%
    \ifnum\x=\y\relax
      \mathbfsbilow{#1}%
    \else
      \mathsfbi{#1}%
    \fi
  \else
    \mathsfbi{#1}%
  \fi
  \endgroup
}
\makeatother

\newcommand{\pD}[2]{\frac{\partial #2}{\partial #1}}

\newcommand{\D}[2]{\frac{{\rm d} #2}{{\rm d} #1}}

\newcommand\bb[1]{\mbox{\boldmath{$#1$}}}
\newcommand\grad{\bb{\nabla}}

\newcommand\btimes{\,\bb{\times}\,}

\newcommand{\msb}[1]{\mathsfbi{#1}}

\newcommand{\mrm}[1]{\mathrm{#1}}

\newcommand{\ex}{\hat{\bb{x}}}
\newcommand{\ey}{\hat{\bb{y}}}

\newcommand{\const}{{\rm const}}

\newcommand{\Br}{B_\mrm{r}}

\newcommand{\pegpp}{{\tt Pegasus++}}

\defcitealias{kunz19}{Kunz \& Squire {\em et al.}~2019}

\title[Triggering tearing with mirrors]{Triggering tearing in a forming current sheet with the mirror instability}

\author[H.W.~Winarto and M.W.~Kunz]%
{Himawan W.~Winarto\thanks{Email address for correspondence: hwinarto@princeton.edu} and Matthew W.~Kunz}

\affiliation{Department of Astrophysical Sciences, Princeton University, Peyton Hall, Princeton, NJ 08544, USA\\[\affilskip]
Princeton Plasma Physics Laboratory, PO Box 451, Princeton, NJ 08543, USA}

\pubyear{2021}
\volume{}
\pagerange{}
\date{\today}
\begin{document}

\maketitle

\begin{abstract}
We study the time-dependent formation and evolution of a current sheet (CS) in a magnetized, collisionless, high-beta plasma using hybrid-kinetic particle-in-cell simulations. An initially tearing-stable Harris sheet is frozen into a persistently driven incompressible flow so that its characteristic thickness gradually decreases in time. As the CS thins, the strength of the reconnecting field increases, and adiabatic invariance in the inflowing fluid elements produces a field-biased pressure anisotropy with excess perpendicular pressure. At large plasma beta, this anisotropy excites the mirror instability, which deforms the reconnecting field on ion-Larmor scales and dramatically reduces the effective thickness of the CS. Tearing modes whose wavelengths are comparable to that of the mirrors then become unstable, triggering reconnection on smaller scales and at earlier times than would have occurred if the thinning CS were to have retained its Harris profile. A novel method for identifying and tracking X-points is introduced, yielding X-point separations that are initially intermediate between the perpendicular and parallel mirror wavelengths in the upstream plasma. These mirror-stimulated tearing modes ultimately grow and merge to produce island widths comparable to the CS thickness, an outcome we verify across a range of CS formation timescales and initial CS widths. Our results may find their most immediate application in the tearing disruption of magnetic folds generated by turbulent dynamo in weakly collisional, high-beta, astrophysical plasmas.
\end{abstract}


\section{Introduction}\label{sec:introduction}

Magnetic reconnection is a fundamental plasma process in which the magnetic-field topology is rapidly rearranged, resulting in the conversion of magnetic energy to plasma energy \citep{zy09,lu16}. Despite its broad relevance to a wide variety of space, astrophysical, and terrestrial plasmas, its detailed physics is most often examined in the somewhat restrictive limit of low plasma beta, $\beta \doteq 8\upi p/B^2 \lesssim 1$ or even ${\ll}1$, where $p$ is the thermal pressure and $B$ is the magnetic-field strength. Given that reconnection plays a central role in powering solar flares and coronal mass ejections, and in degrading energy and particle confinement in tokamak plasmas -- both of which are low-$\beta$ environments of practical importance to humanity -- this focus is perhaps not so surprising. However, nearly half of all the baryonic material in the Universe (often referred to as the `warm-hot intergalactic medium', or WHIM) resides in a hot and dilute plasma state, very likely with $\beta\gg 1$. For example, measurements of Faraday rotation from the intracluster medium (ICM) of the nearby Coma galaxy cluster suggest magnetic-field strengths $B\sim 1$--$5~\mu{\rm G}$ \citep{bonafede10}; given the observed number density $n\sim 3\times(10^{-4}$--$10^{-3})~{\rm cm}^{-3}$ and temperature $T\sim 8~{\rm keV}$, this implies ${\sim}10^{14}~{\rm M}_\odot$ worth of magnetized plasma having $\beta \sim 10^2$. For the ICM in other nearby galaxy clusters, such values of $\beta$ appear to be the norm \citep[e.g.,][]{ct02,govoni17}. It is then of both plasma-physical and astrophysical interest to investigate how magnetic reconnection onsets and proceeds in these more commonplace, high-$\beta$ environments.

There are two immediately complicating factors to such a line of inquiry. The first is that many of these high-$\beta$ astrophysical plasmas, such as the ICM, are also weakly collisional. Taking the Coma cluster again as an example, the Coulomb-collisional mean free path there is $\lambda_{\rm mfp}\sim 10~{\rm kpc}$, roughly $10\%$ of the thermal-pressure scale height. Under such conditions, there is little reason to expect the plasma to be in local thermodynamic equilibrium, particularly so given the turbulent state in which the ICM often finds itself \citep[e.g.,][]{churazov12,hitomi}. Instead, the velocity distribution function of the plasma particles is likely to be biased with respect to the local magnetic-field direction \citep{cgl56,braginskii65}, on account of the plasma's strong magnetization ({\it viz.}, $\rho_i/\lambda_{\rm mfp}\lll 1$, where $\rho_i$ is the Larmor radius of the resident ions; in Coma, $\rho_i \sim 1~{\rm npc}$). Coupled with the high plasma beta, this opens up the potential for kinetic instabilities to feed off departures from velocity-space isotropy (e.g., firehose and mirror instabilities driven by field-aligned pressure anisotropy; \citealt{schekochihin08}) and thereby impact the material properties of the plasma \citepalias{kunz19}.

In the context of magnetic reconnection, the incorporation of pressure anisotropy ({\em viz.}, $p_\perp\ne p_\parallel$, the subscripts denoting the components of the pressure tensor perpendicular and parallel to the magnetic-field direction) has fallen roughly into two categories. The first concerns the linear theory of tearing modes in a pressure-anisotropic environment \citep{cd81,coppi83,cp84,cl85,bc89,shi87,karimabadi04,quest10}, the general finding being that $p_\perp > p_\parallel$ increases tearing growth rates (particularly at smaller scales) while $p_\perp < p_\parallel$ reduces growth rates. Kinetic simulations of reconnection in pressure-anisotropic plasmas support this finding \citep{ambrosiano86,matteini13,gingell15}, with the latter two references drawing attention to the impact of ion-cyclotron instability (for $p_\perp>p_\parallel$) and firehose instability (for $p_\perp<p_\parallel$) on the geometry of the tearing CS. The second concerns pressure anisotropy (particularly in the electrons) that is generated in the reconnecting layer during the nonlinear evolution of the tearing instability, due to particle trapping within the reconnecting layer \citep{egedal13} and/or Fermi acceleration in evolving magnetic islands \citep{schoeffler11}. In both cases -- whether pressure anisotropy is assumed {\em ad hoc} to be present in the initial configuration, or whether it is generated during the reconnection process itself -- the basic lesson is the same: magnetic reconnection is affected by the thermodynamic state of a weakly collisional plasma.

The second complicating factor is the difficulty in formulating a suitable equilibrium state about which to investigate tearing and reconnection. This, of course, holds true regardless of whether $\beta$ is low or high, or whether the plasma is weakly collisional or magnetohydrodynamic (MHD). Indeed, the recent realization that elongated Sweet--Parker-like CSs \citep{parker57,sweet58} are super-Alfv\'{e}nically unstable to a resistive plasmoid instability implies that they can never form in the first place \citep{loureiro07,bhattacharjee09,samtaney09,pv14,tenerani15,lu16}. Instead, one must consider the growth of the tearing instability in a current layer as it is being formed, in order to determine the reconnection onset time and the CS parameters at that moment \citep{ul16}. What high beta and low collisionality bring to the mix are the generation of kinetically unstable pressure anisotropy and the consequent destabilization of the CS on ion-Larmor scales. The argument runs as follows \citep[][]{ak19}.

During the gradual formation of a CS, the strength of the reconnecting magnetic field will increase in the inflowing fluid elements due to flux accumulation. The approximate conservation of adiabatic invariants in the magnetized plasma will then lead to the increase of perpendicular pressure and the decrease of parallel pressure \citep{cgl56}, thus creating pressure anisotropy with $p_\perp > p_\parallel$. At high beta, and in the absence of strong collisional isotropization, the pressure anisotropy will promptly grow large enough to trigger the mirror instability, forming static magnetic mirror-like structures on ion-Larmor-scales that regulate the anisotropy by trapping and scattering particles \citep{kunz14,riquelme15}. These structures influence the CS's stability to tearing by decreasing appreciably the effective thickness of the CS and nonlinearly seeding small-scale tearing modes. The former effect is similar to that caused by small current corrugations in the CS, which \citet{militello2009} showed can boost substantially the value of the tearing-instability parameter. 

The purpose of this paper is then to investigate the early evolution and subsequent disruption of a thinning CS in a high-$\beta$, weakly collisional plasma, while taking into consideration plasma-kinetic effects such as pressure anisotropy and the mirror instability that are self-consistently driven during CS formation. We do so by performing a series of hybrid-kinetic particle-in-cell (PIC) simulations of a thinning CS using the {\tt Pegasus++} code, with the further aim of testing certain predictions of the analytic model of \citet{ak19} for the mirror-instigated tearing of a forming CS.

The paper is organized as follows. Section~\ref{sec:ingredients} provides an exposition of the principal themes of this work, namely, the steady thinning of a forming CS, the attendant adiabatic production of positive pressure anisotropy, and the ensuing destabilization of the sheet by mirror and tearing instabilities. The view of a CS as a dynamically forming structure, rather than a predetermined equilibrium state, is essential to this narrative. Having established these themes, in section~\ref{sec:method} we detail our numerical method for studying CS formation, mirror-stimulated tearing, and the onset of magnetic reconnection. We also describe a novel technique for identifying reconnecting X-points and magnetic islands based on watershed segmentation. The results of our numerical experiments are catalogued in section~\ref{sec:results}. We close in section~\ref{sec:summary} with a recapitulation of our main results, their limitations, and a discussion of their implications for magnetic reconnection in weakly collisional, high-$\beta$ plasmas.

\section{Physical ingredients}\label{sec:ingredients}

We begin our exposition by emphasizing that, in general, any study of the onset problem of magnetic reconnection should take into account effects that transpire and/or accumulate during the formation and gradual thinning of the CS. In the context of this paper, these effects include the adiabatic production of pressure anisotropy, the consequent destabilization of the plasma to the growth of ion-Larmor-scale magnetic mirrors, and the impact of these mirrors on the emergence of tearing modes. These topics are discussed in sections~\ref{sec:paniso}--\ref{sec:mirrorinfest}. But first, we introduce a simple model of CS formation that demonstrates the physics emphasized in this paper and which we ultimately implement in our numerical simulations (\S\ref{sec:csformation}), and then analyze its time-dependent stability to hyper-resistive tearing modes (\S\ref{sec:hyper}).

\subsection{Current-sheet formation}\label{sec:csformation}

Consider a reversing, time-dependent magnetic field with a \citet{harris62} profile without a guide field,
\begin{equation}\label{eqn:CS}
\bb{B}_{\rm r}(t,x) = \Br(t)\tanh\biggl[\frac{x}{a(t)}\biggr]\ey ,    
\end{equation}
which is frozen into an incompressible flow given by
\begin{equation}\label{eqn:flow}
    \bb{u}(t,x,y) = \frac{1}{\rmGamma(t) \tau_{\rm cs}} ( -x\ex + y\ey )  \quad {\rm with} \quad \rmGamma(t) \doteq 1 + \frac{t}{\tau_{\rm cs}} .
\end{equation}
This flow compresses the CS in the $x$ direction and stretches it in the $y$ direction over a characteristic CS formation timescale $\tau_{\rm cs}$, such that the CS maintains its profile but with a decreasing thickness $a(t)=a_0/\rmGamma(t)$ and increasing field strength $\Br(t)=B_{\rm r0}\rmGamma(t)$ \citep[][\S 2]{tolman18}. (Here, and for the remainder of this paper, the `0' subscript denotes an initial value.) The characteristic Alfv\'{e}n Mach number of the flow \eqref{eqn:flow} is
\begin{equation}\label{eqn:Mach}
    M_{\rm A0} \doteq \frac{a_0}{v_{\rm A0}\tau_{\rm cs}},
\end{equation}
where $v_{\rm A0}$ is the initial Alfv\'{e}n speed of the reconnecting field. While the magnetic-field profile $\bb{B}_{\rm r}(t,x)$ has no intrinsic lengthscale in the $y$ direction, any two points initially separated by a distance $L_0\ey$ will move apart and eventually be separated by $L_0\rmGamma(t)\ey$, consistent with the incompressibility of the flow. Accordingly, we define a lengthscale $L(t)=L_0\rmGamma(t)$ and refer to it as the `length' of the CS; while the constant $L_0$ is somewhat arbitrary at this point, it will find a practical definition in our numerical simulations as the initial length of the computational domain in the $y$ direction (see \S\ref{sec:method}).

The thinning of this CS brings with it two important consequences. First, because the aspect ratio $L(t)/a(t)=(L_0/a_0)\rmGamma^2(t)$, the CS will become increasingly susceptible to tearing. Namely, the Harris-sheet tearing instability parameter
\begin{equation}\label{eqn:deltaprime}
    \Delta'(t,N) = \frac{2N}{L(t)} \left[ \frac{1}{N^2} \frac{L^2(t)}{a^2(t)}-1\right] 
\end{equation}
will grow positively for all mode numbers $N\doteq k_{\rm t}(t)L(t)=\const$, where $\bb{k}_{\rm t}(t) = [k_{{\rm t}0}/\rmGamma(t)]\ey$ is the tearing wavenumber made time-dependent due to its advection by the flow \eqref{eqn:flow}. Even for an initially tearing-stable CS satisfying $2\upi a_0/L_0\ge 1$ (for which $\Delta'(0,N)\le 0$ for all $N$), the CS will eventually become unstable as it continually thins. \citet{ul16} considered this time-dependent problem for the onset and evolution of the tearing instability of a thinning CS in the limit of resistive MHD. In section~\ref{sec:hyper}, we adapt their arguments for the case in which the resistive diffusion of the magnetic field is fourth order in space, {\em viz.}~$\eta\nabla^2\bb{B}\rightarrow\eta_4\nabla^4\bb{B}$, relevant to the simulations described in section~\ref{sec:method}.

The second consequence of the thinning of the CS, at least for a sufficiently collisionless plasma, is the adiabatic production of pressure anisotropy in the inflowing fluid elements and, at sufficiently large $\beta$, the eventual triggering of the mirror instability. These events will of course affect any subsequent tearing of the CS -- indeed, this is the entire point of this paper -- but it is instructive to ask first how the CS would evolve without such interference.

\subsection{Hyper-resistive tearing of a thinning CS without mirrors}\label{sec:hyper}

Our first goal is to establish for how long a given CS must thin until tearing modes are able to onset and disrupt its formation. For that, we adapt the arguments given in \citet{ul16} to the case with hyper-resistive tearing, relevant to the simulations described in section~\ref{sec:method} that use fourth-order magnetic dissipation. The idea is as follows: because the instability parameter $\Delta'(t,N)$ (see \eqref{eqn:deltaprime}) increases in time for each unstable tearing mode $N$, the growth rate $\gamma_{\rm t}(t,N)$ of each tearing mode increases as well, with the onset of tearing occurring only once $\gamma_{\rm t}(t,N)\tau_{\rm cs}\gtrsim 1$. A further complication in the case of tearing is that, as the CS lengthens, more and more modes ({\em viz.}, larger values of $N$) become successively unstable, i.e., their $\Delta'(t,N)$ becomes positive. One must then know how $\gamma_{\rm t}$ depends on $N$ to assess whether the tearing is ultimately dominated by a single island or by multiple islands -- a question that depends upon whether $\Delta'$ is small (${\ll}\delta^{-1}_{\rm in}$, where $\delta_{\rm in}$ is the thickness of the resistive inner layer, corresponding to the `constant-$\psi$' or `FKR' regime; \citealt{fkr63}) or large ($\sim\delta^{-1}_{\rm in}$, corresponding to the `Coppi' regime; \citealt{coppi76}). 

For a Harris sheet subject to Ohmic resistivity, the tearing growth rate in the FKR regime satisfies $\gamma_{\rm t} \propto k^{-2/5}_{\rm t}$ for tearing wavenumber $k_{\rm t}$ satisfying $k_{\rm t}a\ll 1$. The fastest-growing FKR mode is then the longest one that fits into the CS (i.e., $N\sim 1$). With hyper-resistivity, however, the growth rate in the FKR regime is roughly independent of $k_{\rm t}$ for $k_{\rm t}a\ll 1$  \citep{hbf13}:
\begin{subequations}\label{eqn:gammat}
\begin{align}
    \gamma_{\rm t} &\sim \frac{v_{\rm A}}{a} S^{-1/3}_a (\Delta' a)^{2/3} (k_{\rm t}a)^{2/3} \label{eqn:gammat_a}\\*
    \mbox{} &\sim \frac{v_{\rm A}}{a} S^{-1/3}_a \left( 1 - k^2_{\rm t} a^2\right)^{2/3} , \quad{\rm where}\quad S_a \doteq \frac{a^3v_{\rm A}}{\eta_4} \label{eqn:gammat_b}
\end{align}
\end{subequations}
is the hyper-resistive Lundquist number and $v_{\rm A}$ is the Alfv\'{e}n speed associated with the reconnecting field $B_{\rm r}$. In the corresponding Coppi regime, \citet{hbf13} find slower growth with $\gamma_{\rm t}\propto k^{4/5}_{\rm t}$ for $k_{\rm t}a\lesssim S^{-1/6}_a$ (see their figure~1). As a result, the fastest-growing unstable mode will always be an FKR-like mode with $\gamma_{\rm t} \sim (v_{\rm A}/a)S^{-1/3}_a$. 

With all such modes growing at approximately the same rate, we argue that it will be the mode that first becomes unstable as the current sheet thins, {\em viz.}~$N\sim 1$, that will ultimately win out (it having a head start over the others). Demanding that its growth rate satisfy $\gamma_{\rm t}\tau_{\rm cs}\gtrsim 1$ requires that $a \lesssim v_{\rm A} \tau_{\rm cs} S^{-1/3}_a \ll L$. Using our CS model with $L(t)/L_0=a_0/a(t)=v_{\rm A}(t)/v_{\rm A0}=1+t/\tau_{\rm cs}$ (see \S\ref{sec:csformation}) then specifies constraints on the critical time $t_{\rm cr}$ at which $\gamma_{\rm t}\tau_{\rm cs}\gtrsim 1$:\footnote{If the right-hand side of \eqref{eqn:tcrit} were to become ${\gtrsim}S_{a0}^{4/21} (a_0/L_0)^{8/7}$ before \eqref{eqn:tcrit} is satisfied, then the $N\sim 1$ mode would enter the slower-growing Coppi regime before its FKR growth rate satisfies $\gamma_{\rm t}\tau_{\rm cs}\gtrsim 1$. In this case, equation \eqref{eqn:tcrit} would pertain only to those unstable modes whose mode numbers are greater than or equal to that of the mode `transitional' between the FKR and Coppi regimes, {\em viz.}~$N_{\rm tr} \sim (L/a) S^{-1/6}_a$. Evaluating the latter at $t=t_{\rm cr}$, we find that all unstable modes whose $N \ge N_{\rm tr}(t_{\rm cr}) \sim (L_0/a_0) S^{1/8}_{a0} M^{7/8}_{\rm A0}$ attain growth rates satisfying $\gamma_{\rm t}\tau_{\rm cs}\gtrsim 1$ at roughly the same time. None of these details affect the estimate implied by \eqref{eqn:tcrit} for $t_{\rm cr}$.}
\begin{equation}\label{eqn:tcrit}
    \left(1+\frac{t_{\rm cr}}{\tau_{\rm cs}}\right)^{2/3} \frac{a_0}{L_0} \ll S_{a0}^{1/3} M_{\rm A0} \lesssim \left( 1 + \frac{t_{\rm cr}}{\tau_{\rm cs}} \right)^{8/3} .
\end{equation}
With a view towards the results presented in section~\ref{sec:results}, we note that our numerical simulations generally have $S_{a0} \sim 10^6$ and $M^{-1}_{\rm A0} \in[1,16]$; the inequality \eqref{eqn:tcrit} then implies values of $t_{\rm cr}/\tau_{\rm cs}$ that exceed unity by a factor of a few. In other words, for a thinning Harris CS without the production of pressure anisotropy and consequent excitation of mirror modes, we expect linear hyper-resistive tearing of our forming CS to onset at $N\sim 1$ after a few $\tau_{\rm cs}$. It is important to bear this point in mind for what follows.

\subsection{Production and destabilization of pressure anisotropy}\label{sec:paniso}

We now demonstrate that the arguments given in section~\ref{sec:hyper} for the onset of tearing in a thinning CS are likely to be irrelevant in a collisionless, high-beta plasma. 

Consider a flux-frozen fluid element initially located at $x=\xi_0$ and moving towards $x=0$ in the velocity field \eqref{eqn:flow}. It is straightforward to show that, as the CS thins, the magnetic-field strength as seen by this element steadily increases as $B_{\rm r0}\tanh(\xi_0/a_0)\rmGamma(t)$. If the first and second adiabatic invariants of the plasma are approximately conserved during this thinning, then the components of the pressure tensor parallel ($p_\parallel$) and perpendicular ($p_\perp$) to the local magnetic field become unequal, with the latter outpacing the former. Namely, if the plasma is initially pressure-isotropic, then the pressure anisotropy $\Delta_p \doteq p_\perp/p_\parallel - 1$ in the fluid element should evolve according to
\begin{equation}\label{eqn:Deltap}
    \Delta_p(t,\xi(t)) = [\rmGamma(t)]^3 - 1 \approx \frac{3t}{\tau_{\rm cs}} ,
\end{equation}
the approximation being accurate for $t/\tau_{\rm cs}\ll 1$. Thus, pressure anisotropy increases in all fluid elements at the same rate.

In a plasma whose collision timescale is much longer than the CS formation timescale, the pressure anisotropy will continue to grow in time following \eqref{eqn:Deltap} until one of two things occurs: either (i) the tearing instability onsets and disrupts the steady thinning of the CS, or (ii) the pressure anisotropy grows large enough to surpass the mirror instability threshold,
\begin{equation}\label{eqn:Lambdam}
    \Lambda_\mrm{m}\doteq\Delta_p-\frac{1}{\beta_\perp}>0 ,
\end{equation}
where $\beta_\perp \doteq 8\upi p_\perp / B^2$. With $\beta_\perp(t,\xi(t)) = \beta_\perp(0,\xi_0)/\rmGamma(t)$, the latter scenario is possible in any inflowing fluid element so long as $\beta_\perp(0,\xi_0) > 4^{1/3}/3 \simeq 0.53$. Let us assume for the moment that case (ii) happens first, likely a good assumption when $\beta_\perp(0,\xi_0)\gg 1$, since in this situation the plasma becomes mirror-unstable ({\em viz.}~$\Lambda_{\rm m}>0$) after a small fraction of the CS formation time. Namely, setting \eqref{eqn:Deltap} equal to $1/\beta_\perp(t,\xi(t))$ and Taylor-expanding the result in $t/\tau_{\rm cs}\sim 1/\beta_\perp(0,\xi_0)\ll 1$, we find that the plasma becomes mirror-unstable at a time $t_{\rm m}$ that satisfies
\begin{equation}
    \frac{t_{\rm m}}{\tau_{\rm cs}} \approx \frac{1}{3\beta_\perp(0,\xi_0)} \ll 1.
\end{equation}
In the next two subsections, we use this estimate to predict the evolution of mirror fluctuations, including when they should regulate the pressure anisotropy and what the consequences are for tearing in a mirror-infested sheet.

\subsection{Mirror instability in a forming CS}\label{sec:mirror}

For a uniform plasma with $\Lambda_{\rm m}>0$ threaded by a static, uniform magnetic field, the maximum growth rate of the mirror instability is given by $\gamma_{\rm m}\sim \Omega_i \Lambda^2_{\rm m}$, where $\Omega_i$ is the ion-Larmor frequency \citep{hellinger07}. In the asymptotic limit $\beta_\perp\Lambda_{\rm m}\ll 1$, the field-parallel and field-perpendicular wavenumbers at which this growth occurs satisfy $k_{\parallel,{\rm m}}\rho_i \sim (k_{\perp,{\rm m}}\rho_i)^2\sim \Lambda_{\rm m}$, where $\rho_i$ is the ion-Larmor radius. Two things complicate the application of these formulae to our forming CS. First, the reconnecting field \eqref{eqn:CS} is non-uniform, with a null line occurring at the centre of the CS. This means that both $\Omega^{-1}_i$ and $\rho_i$ are smaller away from the neutral line, with ions unable to execute Larmor motion for $|x|\lesssim(\rho_i a)^{1/2}$ \citep{parker57,dobrowolny68,cp84}. As a result, mirror modes will grow faster and on smaller physical scales away from the CS ($|x|\gtrsim a$), where the plasma is better magnetized.

Secondly, the mirror growth rate and wavenumbers are made time-dependent by the increasingly positive pressure anisotropy, and mirror modes can only emerge in the magnetized regions of the CS if their growth rate is much larger than the rate of CS formation, i.e.~$\gamma_{\rm m}(t) \tau_{\rm cs}\gg 1$. With $\gamma_{\rm m}(t) \sim \Omega_i \Lambda^2_{\rm m}(t)$, this means that the instability parameter $\Lambda_{\rm m}$ must reach a value ${\gg}(\Omega_i \tau_{\rm cs})^{-1/2}$ before the mirrors can grow fast enough to outpace the more leisurely production of pressure anisotropy. Thereafter, $\Lambda_{\rm m}$ will decrease as the plasma converts its free energy into magnetic fluctuations. We therefore anticipate a maximal value of $\Lambda_{\rm m}$ given by
\begin{equation}\label{eqn:lambdamax}
    \Lambda_{\rm m,max} = C_{\rm m} (\Omega_i\tau_{\rm cs})^{-1/2} ,
\end{equation}
where $C_{\rm m}\gg 1$ is some constant (determined by the numerical simulations in section~\ref{sec:results} to be ${\approx}14$).\footnote{A similar scaling was found empirically by \citet{kunz14} in their simulations of positive pressure anisotropy driven by a linear shear flow and of the resultant mirror instability.} By Taylor-expanding $\Lambda_{\rm m}(t)$ about $t=t_{\rm m}$, we find that \eqref{eqn:lambdamax} is attained after a time $t_{\rm m,reg}$ that satisfies
\begin{equation}
    \frac{t_{\rm m,reg}-t_{\rm m}}{\tau_{\rm cs}} \approx \frac{C_{\rm m}}{3} (\Omega_i\tau_{\rm cs})^{-1/2},
\end{equation}
beyond which the pressure anisotropy is regulated towards the mirror-instability threshold, $\Lambda_{\rm m}=0$. Due to the mirrors' super-exponential growth, this regulation will occur very rapidly after $t_{\rm m,reg}$.

As $\Lambda_{\rm m}$ tends towards zero, the transfer of free energy from pressure anisotropy to mirror fluctuations drives the fluctuation amplitude $\delta B/B$ to a value ${\sim}\Lambda^{1/2}_{\rm m,max}$ \citep{kunz14}. But as the CS continues to thin,  $\Delta_p$ continues to be driven positively, and the mirror fluctuations must grow in amplitude to maintain a marginally unstable plasma. For a linear-in-time drive, they do so secularly with $\delta B^2 \propto t^{4/3}$ \citep{schekochihin08,kunz14,rincon15}, as an increasing fraction of large-pitch-angle particles become trapped in the deepening magnetic wells. If tearing does not intercede and disrupt their evolution, these mirrors would grow in amplitude all the way to $\delta B/B \sim 0.3$, independent of $\Lambda_{\rm m,max}$, after which the ions pitch-angle scatter off sharp ion-Larmor-scale bends in the mirroring field and saturate the instability. This scattering breaks adiabatic invariance and thereby maintains marginal stability by severing the link between $\Delta_p$ and changes in $B$ \citep{kunz14,riquelme15}. In the context of our forming CS, reaching this saturated state would take a time ${\sim}\tau_{\rm cs}$.

\subsection{Hyper-resistive tearing of a mirror-infested CS}\label{sec:mirrorinfest}

The emergence of mirrors in the forming CS has two impacts on any subsequent tearing. First, by appreciably wrinkling the reconnecting field on kinetic scales, nonlinear mirrors cause $\Delta'$ to depart from that corresponding to an undisturbed Harris sheet, equation \eqref{eqn:deltaprime}. Proposing a simple model for the magnetic profile of a mirror-infested CS, \citet{ak19} calculated the resulting $\Delta'(k_{\rm t})$ and showed that it becomes positive (and thus the CS becomes tearing-unstable) at wavenumbers $k_{\rm t}$ up to that corresponding to the inverse location of the innermost magnetic mirror (which effectively replaces $a$ as the characteristic thickness of the CS). Those authors then argued that the location of the innermost magnetic mirror is set by the perpendicular wavenumber of that mirror, using $k_{\perp,{\rm m}}\rho_i \sim \Lambda^{1/2}_{\rm m,max}$ with $\Lambda_{\rm m,max}$ given by \eqref{eqn:lambdamax} and $\rho_i$ and $\Omega_i$ taking on their local values. In the case of a very weak guide field, and accounting for the Lagrangian compression during CS formation, this corresponds to a distance $x_{\rm m}$ that satisfies
\begin{equation}\label{eqn:ym}
    \frac{x_{\rm m}}{a(t)} \sim \biggl(\frac{\rho_{i0}}{a_0}\biggr)^{4/7} (\Omega_{i0}\tau_{\rm cs})^{1/7}
\end{equation} 
(see their equation (4.11)). For $a_0/\rho_{i0}\gg (\Omega_{i0}\tau_{\rm cs})^{1/4}$, this dramatically extends the range of tearing-unstable wavenumbers (from $k_{\rm t}\sim 1/a$ up to ${\sim}1/x_{\rm m}$). It also expedites the onset of reconnection by boosting tearing growth rates, since the effective $\Delta' \sim 1/x_{\rm m}$ is independent of $k_{\rm t}$ and so equation \eqref{eqn:gammat_a} implies $\gamma_{\rm t} \propto k^{2/3}_{\rm t}$. Secondly, the perturbations to the reconnecting field caused by the mirror instability will nonlinearly seed tearing modes with $k_{\rm t}\gtrsim k_{\parallel,\rm m}$, giving them a head start over any traditional tearing modes of the thinning Harris sheet (if they are indeed unstable). As a result, mirror-stimulated tearing should onset at a wavenumber $k_{\rm t}$ satisfying $x^{-1}_{\rm m} \gtrsim k_{\rm t} \gtrsim k_{\parallel,{\rm m}}$.

Taken together, these two effects suggest that the onset of reconnection in an evolving CS, driven by mirror-stimulated tearing modes, likely occurs earlier and at smaller scales than it would have without the mirrors. More quantitatively, if pre-tearing mirrors were to wrinkle the CS and effectively reduce the CS thickness by a factor ${\sim}(x_{\rm m}/a)$ (see \eqref{eqn:ym}), then the right-hand side of \eqref{eqn:tcrit} would acquire a multiplicative factor ${\sim}(a/x_{\rm m})^{4/3} \sim (a_0/\rho_{i0})^{16/21}(\Omega_{i0}\tau_{\rm cs})^{-4/21}\gg 1$, thereby  reducing the critical time significantly.

The remainder of the paper is dedicated to testing these ideas using numerical simulations of CS formation and reconnection onset.

\section{Method of solution}\label{sec:method}

\subsection{Hybrid-kinetic treatment of a thinning CS}\label{sec:expbox}

For our numerical simulations, we adopt the hybrid-kinetic approximation, in which a non-relativistic, quasi-neutral, collisionless plasma is modeled by kinetic ions (mass $m_i$, charge $e$) that interact with a massless electron fluid via an electric field $\bb{E}$ given by the following generalized Ohm's law:
\begin{equation}\label{eqn:Efield}
    \bb{E} = -\frac{\bb{u}}{c}\btimes\bb{B} - \frac{T_e}{en} \grad n + (\grad\btimes\bb{B})\btimes\frac{\bb{B}}{4\upi en} .
\end{equation}
Our notation is standard: $n$ and $\bb{u}$ are the ion number density and bulk-flow velocity, respectively; $T_e$ is the (assumed constant and isotropic) temperature of the electron fluid; $c$ is the speed of light; and the magnetic field $\bb{B}$ satisfies Faraday's law of induction,
\begin{equation}\label{eqn:induction}
    \pD{t}{\bb{B}} = -c\grad\btimes\bb{E} .
\end{equation}
The ions are treated using the PIC method, in which they are represented by finite-sized macro-particles whose positions $\bb{r}_p$ and velocities $\bb{v}_p$ are governed by the characteristic equations
\begin{align}\label{eqn:ions}
    \D{t}{\bb{r}_p} &= \bb{v}_p , \\*
    \D{t}{\bb{v}_p} &= \frac{e}{m_i} \left[ \bb{E}(t,\bb{r}_p) + \frac{\bb{v}_p}{c}\btimes\bb{B}(t,\bb{r}_p)\right] .\label{eqn:ionv}
\end{align}
These equations are solved using the second-order--accurate code \pegpp~(Arzamasskiy et al., in prep); we refer the reader to \cite{kunz14_pegasus} for algorithmic details.

To model the thinning of the CS, we impose an externally driven, immutable flow that incompressibly expands the plasma along the sheet direction while contracting it in the perpendicular direction. To avoid having the box itself deform, we perform the simulation in the frame of the flow by applying a continuous coordinate transformation to the equations during each time step of the integration. To do so, we introduce the time-dependent Jacobian transformation matrix $\msb{\rmLambda}(t)\doteq\partial\bb{r}/\partial\bb{r}'$ tying the lab frame $(t,\bb{r},\bb{v})$ to the comoving frame $(t'=t,\bb{r}'=\msb{\rmLambda}^{-1}\bb{r},\bb{v}'=\msb{\rmLambda}^{-1}\bb{v})$, and specify its form via
\begin{equation}
    \msb{\rmLambda}(t) = \rmGamma^{-1}(t) \ex\ex + \rmGamma(t) \ey\ey,\quad{\rm with}\quad \rmGamma(t) \doteq 1 + \frac{t}{\tau_{\rm cs}} .
\end{equation}
This transformation has the consequence that the strength of the reconnecting field increases linearly in time as the current-sheet thickness $a(t)\propto 1/\rmGamma(t)$, as in section~\ref{sec:csformation}. Note that $\lambda \doteq \det\msb{\rmLambda}=1$ at all times, thereby preserving the area of each cell and thus their density. In this comoving (primed) frame, equations \eqref{eqn:Efield}--\eqref{eqn:ionv}  become, respectively,
\begin{align}
    \bb{E}' &= - \frac{\bb{u}'}{c}\btimes\bb{B}' - \frac{T_e}{en'}\grad' n' + (\grad'\btimes\bb{B}')\btimes\frac{\msb{\rmLambda}^2\bb{B}'}{4\upi e n'\lambda}, \\*
    \pD{t'}{\bb{B}'} &= -c\grad'\btimes\bb{E}' - \eta_4 \nabla'^4 \bb{B}', \label{eqn:bprime}\\*
    \D{t'}{\bb{r}'_p} &= \bb{v}'_p , \\*
    \D{t'}{\bb{v}'_p} &= \frac{e}{m_i} \msb{\rmLambda}^{-2} \biggl[ \bb{E}'(t',\bb{r}'_p) + \frac{\bb{v}'_p}{c}\btimes\bb{B}'(t',\bb{r}'_p) \biggr] - 2\msb{\rmLambda}^{-1} \D{t'}{\msb{\rmLambda}}\, \bb{v}'_p ,\label{eqn:ionvp}
\end{align}
where $\bb{E}'=\msb{\rmLambda}\bb{E}$, $\bb{B}'=\lambda\msb{\rmLambda}^{-1}\bb{B}$, $n'\doteq \lambda n$, and $\bb{u}'\doteq\msb{\rmLambda}^{-1}\bb{u}$ \citep[see][appendix A]{ht05}. To facilitate the reconnection of magnetic-field lines, we have appended to \eqref{eqn:bprime} a fourth-order hyper-resistive diffusion with constant coefficient $\eta_4$ (discussed further below). This is the set of equations solved by \pegpp; the final (velocity-dependent) term in equation \eqref{eqn:ionvp} is straightforwardly incorporated into the semi-implicit Boris algorithm for solving particle trajectories alongside the $\bb{v}'_p\btimes\bb{B}'$ rotation. Quantities in the lab frame are easily obtained {\it ex post facto}.

Before proceeding any further, we pause here for a moment to explain our adoption of a hybrid-kinetic model with a fourth-order hyper-resistivity, rather than an alternative approach in which the electron kinetics play a role and the reconnection is facilitated by electron inertia and pressure-tensor effects. In some ways, our use of hybrid-kinetics is of a purely pragmatic nature: our focus is on the macroscale evolution of a forming CS, the generation of pressure anisotropy in the inflowing fluid elements, and the effect of emergent mirror modes on the stability of the thinning sheet. All of this physics is captured within hybrid-kinetics, which has the added bonus of being significantly cheaper numerically than a fully kinetic approach (especially at high $\beta$ with large scale separations). Moreover, hyper-resistivity is often used to mimic the role of anomalous electron viscosity in facilitating reconnection, and yields tearing growth rates in the FKR regime that are approximately independent of $k_{\rm t}$ for a Harris sheet (recall \eqref{eqn:gammat_b}), just as in the fully collisionless case \citep[e.g.,][]{dl77,karimabadi05,fp04,fp07}. 

Practical considerations aside, however, there is some astrophysical justification for resolving the ion-Larmor scale while breaking the frozen-in condition through a resistive term at sub-$\rho_i$ scales. In the hot and dilute ICM, $\rho_i \sim 1~{\rm npc}$ is far below any meso- or macro-scale (including the Coulomb mean free path) but is only a factor of a few larger than the Ohmic-resistive scale (assuming Coulomb collisions; \citealt{sc06}). Such a scale hierarchy places the ICM (and our simulations) in the `semi-collisional' regime specified by the ordering $d_e \ll \delta_{\rm in}\ll \rho_i \ll a$ where $d_e$ denotes the electron skin depth, in which MHD is no longer a sufficient description but the frozen-flux constraint is broken by resistivity rather than electron inertia \citep[][although their focus was on the low-beta limit]{ckh86,bl18}. All this is to say that our hybrid-kinetic approach is not without physical utility or direct application to actual systems.

Of greater potential consequence is our neglect of electron pressure anisotropy, which could affect the reconnection dynamics in a number of ways. First, both in observations of reconnection in the Earth's magnetotail \citep[e.g.,][]{oieroset02} and in previous theoretical work on collisionless reconnection \citep{egedal13}, pressure anisotropy in the electron species (with $p_\parallel > p_\perp$) has been found to influence the inner diffusive layer of the CS. The idea is that adiabatic trapping of the electrons by magnetic mirrors and parallel electric fields within the reconnection region cause strong parallel heating of the electrons, producing pressure anisotropy that alters the reconnection geometry by driving electron currents in extended layers. Hybrid simulations that incorporate electron pressure anisotropy have shown the formation of more elongated CSs near X-points when compared to those found when adopting a pressure-isotropic closure, although the reconnection rate in both cases was found to be similar \citep{le16}. Because we are considering the case with zero guide field, we do not anticipate such effects to play a role; without a guide field, the electrons are not adiabatically trapped in the inner layer and so can stream freely across the sheet and phase mix with other electrons \citep{le13}. They can, however, develop significant agyrotropy in their pressure tensor as they partially demagnetize in the electron diffusion layer and bounce in the field-reversal region \citep{hesse01}. In collisionless reconnection, this agyrotropy contributes to the reconnecting electric field \citep{vasyliunas75,hesse11} and can be used as a proxy for identifying separatrices and X-points \citep{scudder08,swisdak16}. Another way in which pressure anisotropy can develop during reconnection is due to Fermi acceleration within contracting islands, which increases the parallel pressure and, at sufficiently high plasma $\beta$, becomes regulated by the firehose instability \citep{schoeffler11}. However, all of these effects -- particle trapping within the reconnecting layer and Fermi acceleration during island growth and merging -- feature in CSs within which tearing has already onset and gone non-linear. In the context of this paper, the more relevant missing physics is electron pressure anisotropy generated during the CS formation itself, in a manner analogous to that discussed in section~\ref{sec:paniso} for the ions. In this case, the electron pressure anisotropy could contribute to destabilizing the forming CS at high $\beta$, either by adding to the total pressure anisotropy that factors into the mirror instability threshold \citep{bc82,hellinger07,hs18} or by triggering its own kinetic instabilities (e.g., electron whistler instability; \citealt{kp66,gw96,riquelme16}). All of these effects could be taken into consideration in future work by adopting a fully kinetic approach.

\subsection{Initial and boundary conditions}

For all of our simulations, we construct a doubly periodic, two-dimensional (2D) domain with initial size $L_x\times L_y$, in which we initialize a magnetic field having a double Harris-sheet profile with no guide field,
\begin{align}
    \bb{B}_{\rm r}(t=0) = B_{\rm r0} \biggl[ \tanh\biggl(\frac{x - x_{\rm cs, 1}}{a_0}\biggr) - \tanh\biggl(\frac{x - x_{\rm cs, 2}}{a_0}\biggr) - 1 \biggr] \ey .
\end{align}
The locations of the two CSs are set to be $x_{\rm cs, 1}\doteq L_x/4$ and $x_{\rm cs, 2}\doteq 3L_x/4$. As we vary the initial current-sheet thickness $a_0$ in our parameter study, we keep $L_x/a_0=48$, a value large enough to minimize interactions between the two sheets.

Using this numerical setup eliminates some complications that can arise from adopting the more common (and, arguably, more appropriate) open boundary conditions. In addition to the challenge of runaway particles in the open boundary setup, it is difficult to choose an appropriate particle distribution function in the upstream. On the other hand, by disallowing reconnected magnetic flux from leaving the domain, periodic boundaries lead to unphysical behaviour at long times. Fortunately, with our focus being solely on the impact of pressure anisotropy and mirrors on the onset of tearing modes, and not on the long-time evolution of reconnection, this should not pose any problem. Having two CSs in each simulation has the added benefit of reducing statistical noise when computing mean quantities.

We normalize the magnetic field to the initial strength of the reconnecting field far away from the CS, $B_{\rm r0}$, and all velocities to the initial Alfv\'en velocity, $v_{\rm A0} \doteq B_{\rm r0}/ \sqrt{4\pi m_i \langle n_{i0} \rangle}$, where the initial ion density is averaged over the entire simulation domain and set equal to unity. The simulation time is normalized to the initial ion gyrofrequency, $\Omega_{i0} \doteq e B_{\rm r0}/ m_i c$, and all lengthscales are normalized to the initial ion skin depth, $d_{i0} \doteq v_{\rm A0}/\Omega_{i0}$. For our fiducial case with initial sheet width $a_0 = 125$, we choose $L_x \times L_y = 6000 \times 1500$ with $N_x \times N_y = 2688 \times 672$ cells. The hyper-resistivity is set so that $L^3_y v_{\rm A0}/\eta_4 = 5.625 \times 10^{8}$. The current-sheet length is chosen such that the smallest available parallel wavenumber $k_{y,{\rm min}} \doteq 2\upi / L_y$ is just barely unstable to the tearing instability discussed in section~\ref{sec:hyper}, that is, $k_{y,{\rm min}}a_0 \approx 0.5$. Additional simulations using a wider sheet with $a_0 = 250$ and $L_x=12000$, for which $\Delta'(k_{y,{\rm min}}a_0)<0$, are also presented for comparison. We vary $\tau_{\rm cs} \in [1.25, 2.5, 5, 10, 20] \times 10^2$, which implies characteristic inflow velocities $a_0/\tau_{\rm cs}$ that are sub-Alfv\'enic for all but the fastest compression ($M_{\rm A0}\lesssim 1$).

Initially, we draw $N = 10^4 N_x N_y$ ion macro-particles from a stationary Maxwell distribution corresponding to an initial ion temperature $T_{i0}$ and distribute them spatially to satisfy pressure balance within the double-Harris sheet. The (massless, fluid) electrons are set to be in thermal equilibrium with the ions, $T_e/T_{i0} = 1$. We initialize the ions to be pressure-isotropic with $\beta_{\parallel,i0} = \beta_{\perp,i0} = \beta_{i0}=100$; the initial ion-Larmor radius in code units is then given by $\rho_{i0} \doteq \beta_{i0}^{1/2}=10$. For these physical and numerical parameters, the ion-Larmor scale and the thickness of the diffusive inner layer are resolved at all times. Note, however, that neither the production of ion pressure anisotropy and its regulation by the mirror instability, nor the evolution of the hyper-resistive tearing modes triggered by them within the forming CS, require the resolution of the ion skin depth.

\subsection{Watershed segmentation: identifying magnetic islands and X-points}\label{sec:watershed}

Several useful diagnostics for quantifying the onset of reconnection require the identification and subsequent tracking of the X-points that form along the neutral line of the CS. In 2D reconnection, these points may be identified by tracing isocontours of the magnetic flux function $\psi$ or, equivalently, of the plane-perpendicular component of the magnetic vector potential. In this case, X-point locations are defined as the saddle points of $\psi$, where the first-order derivatives vanish and the second-order derivatives in two orthogonal directions have opposing signs. The latter can be quantified by computing the Hessian matrix of $\psi(x,y)$, defined as $H_\psi(x,y)$, and looking for cells with negative determinants \citep[e.g.,][]{servidio09}. This saddle-point identification method relies heavily on the ability to perform accurate finite differences at the grid level; this may be difficult in data from PIC simulations due to the grid-scale noise coming from the (second-order-accurate) deposition of the ion density and momentum on the grid using a finite number of simulation particles. Typically, this `PIC noise' masks the precise locations of zero crossings in the first-order-derivatives test \citep[e.g.,][]{haggerty17}.

To circumvent this issue, we have tried a different method discussed in \citet{vladimir13} that takes three differently sized loops around each cell and considers the cell to contain an X-point if the values of $\psi(x,y)$ rise above twice and fall below twice of the tested cell's value on each loop. Unfortunately, we have also found this procedure to be sensitive to fluctuations at the grid scale from PIC noise, resulting in spuriously identified X-points. 

We have instead developed a novel method to locate X-points based on the watershed segmentation algorithm \citep{bm18}. This algorithm has been extensively studied and utilized to separate out regions of a scalar field \citep[e.g.,][]{mangan1999}. It treats the value of the field at any given point as though it were the topographic height of a relief and segments the region by its flood basins around local minima (hence the name of the algorithm).

For X-point identification, $\psi(x,y)$ acts as the scalar variable of interest. Depending on the directional change of the reconnecting field, it might be necessary to consider $-\psi(x,y)$ instead, such that locations of the O-points on the neutral line can be the basins' local minima. Following application of the algorithm, each magnetic island behaves similarly to the rainfall basin that surrounds an O-point. Preprocessing by applying a low-pass filter on $\psi(x,y)$ can further eliminate any erroneously segmented regions on sub-$\rho_i$ scales. We found that a Gaussian filter with a radius of size $\rho_{i0}$ works best to minimize the mis-identification of basins. Compared to the other two methods discussed above, this method is much more robust to the grid-scale noise coming from the PIC deposition scheme.\footnote{As this paper was being completed, we became aware of a similar approach to X-point detection proposed and analyzed by \citet{banesh20}, which treats $\psi(x,y)$ as a topological scalar field and implements a contour-tree based segmentation algorithm to identify X-point locations. The goal there, as well, was to reduce the complicating effects of PIC noise on X-point identification.}

Given the watershed algorithm properties, it is possible that some of the segmented regions do not contain any magnetic island. These regions are typically the result of spuriously identified valleys located away from the neutral line and can easily be discarded. For two neighbouring regions that contain magnetic islands, an X-point should exist on their boundary and correspond to the local minimum value along that boundary. In addition, locations of O-points can be easily identified as the local minimum value of the flux function within the regions containing a magnetic island.

\begin{figure}
\centering
\includegraphics[width=\textwidth]{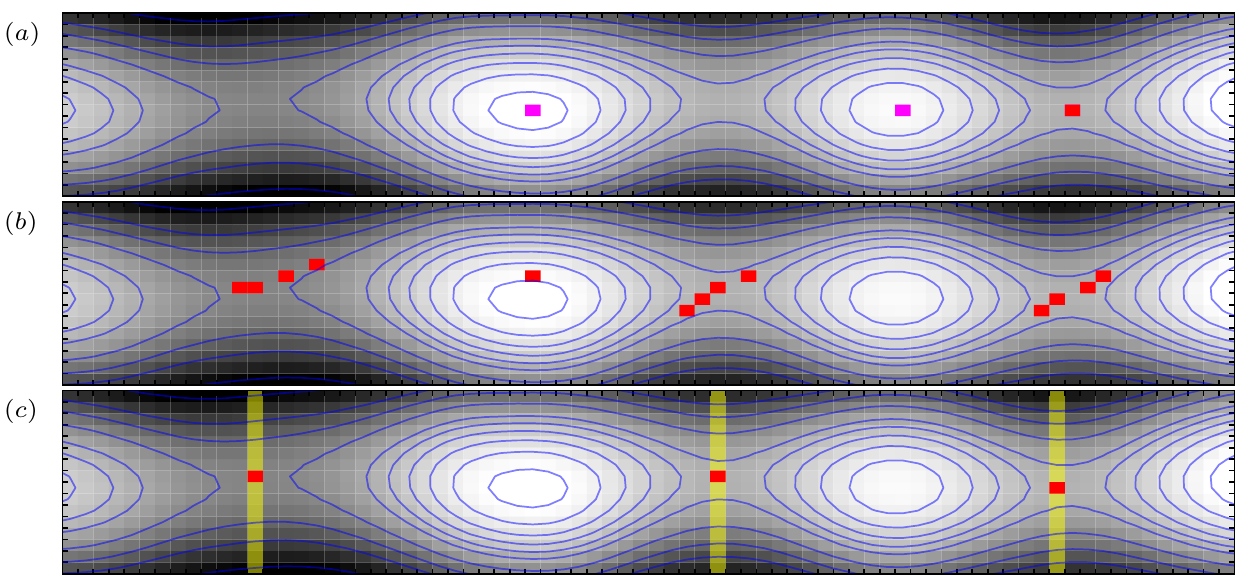}
\caption{Comparison of X-point detection methods for a given flux function $\psi(x,y)$ taken from one of our simulations: ($a$) simple saddle-point, ($b$) loop comparison, and ($c$) watershed segmentation. Values of $\psi(x,y)$ are represented as the grayscale shading and the interpolated blue contours. Cell locations determined to contain X-points are shaded in red, while the purple shaded cells in panel ($a$) show locations of other zero crossings of the first-order derivative. Yellow shaded cells in panel ($c$) show the boundary cells obtained from watershed segmentation.}\label{fig:xmethods}
\end{figure}

In \cref{fig:xmethods} we present a direct comparison between these three algorithms for detecting X-point locations. We select a small region centred about the neutral line from our fiducial simulation (see \cref{fig:cont}($c$)) that clearly exhibits three X-points. Panel ($a$) demonstrates that the simple saddle-point method tends to under-count X-points, identifying only one out of the three X-point locations. The reason for this is that the constraints on the first-order derivative  must be simultaneously satisfied for gradients in both the $x$ and $y$ directions. In contrast, the loop comparison algorithm shown in panel ($b$) tends to over-count, or even mistakenly identify, X-points. Cells satisfying the test conditions can also be disconnected, complicating the interpretation of identified X-point locations. The result from our proposed watershed segmentation algorithm is shown in panel ($c$), and demonstrates a robust detection, identifying all X-point locations.

\section{Results}\label{sec:results}

In this section we present the results from our simulations, in which two CSs that initially have Harris-sheet profiles gradually thin and lengthen according to the coordinate transformation discussed in section~\ref{sec:expbox}. Our results are organized into three subsections: the first focuses on what we consider to be our fiducial run ($a_0=125$, $\tau_{\rm cs}=1000$), while the other two concern variations in the compression time $\tau_{\rm cs}$ and the initial CS width $a_0$. In general, we observe that as the strength of the  reconnecting field increases linearly, $B_{\rm r}(t) = \rmGamma(t)$, pressure anisotropy builds up and eventually triggers the mirror instability. Mirror fluctuations then wrinkle the CSs, changing their magnetic profiles and seeding small-scale tearing modes that would otherwise be stable. These mirror-triggered tearing modes first grow exponentially and then secularly, ultimately spawning multiple islands that merge and grow to become comparable to the instantaneous CS thickness. At that point, which is typically on the order of the CS formation time $\tau_{\rm cs}$, we terminate the simulations and consider reconnection of the CS to have fully onset.

Given the magnetic geometry in our simulations, we focus mainly on four separate regions in the domain whose widths are comparable to $a_0/\rmGamma(t)$ and whose lengths span the full $L_y$. Two of these regions, centred about $x=x_{\rm cs,1}$ and $x_{\rm cs,2}$, represent the dynamics occurring inside each of the two CSs; any quantity averaged over these two regions is referred to with the label \texttt{CS}. The two other regions are centred about $x=0$ and $x=L_x/2$, in between the CSs; any quantity averaged over these two regions is referred to with the label \texttt{Bulk}. All mean quantities are denoted by $\left\langle\, \dots \right \rangle$ unless otherwise noted.

\subsection{Fiducial run}

\begin{figure}
\centering
\includegraphics[width=\textwidth]{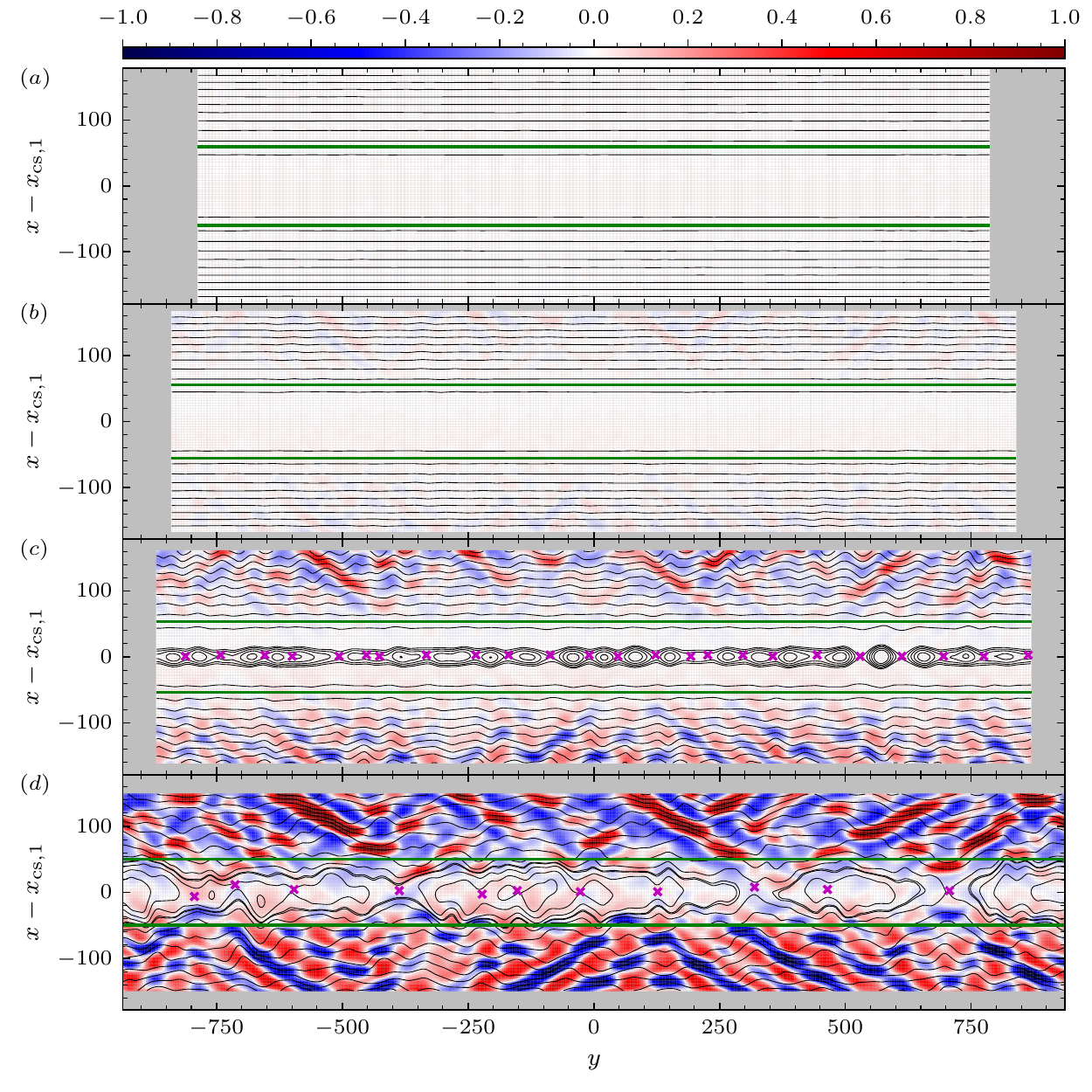}
\caption{Time slices centred about the CS at $x=x_{\rm cs,1}$ from our fiducial simulation, showing properties of the magnetic field at $t = 50$ (panel $a$), 120 (panel $b$), 160 (panel $c$), and 250 (panel $d$). All quantities have been transformed back into the stationary lab frame; note the geometric thinning and lengthening of the CS. The size of an undisturbed Harris sheet of width $a(t)=a_0/\rmGamma(t)$ is marked for reference by the green lines. The quantity $\delta B_y \doteq B_y - \rmGamma(t)\tanh[x/a(t)]$ is represented by the colour contours, with the overlaid black lines tracing levels of the flux function $\psi$. In the bottom two panels, locations of the inferred X-points, found using the method discussed in section~\ref{sec:watershed}, are marked by the purple X's and surrounded by finer levels of the flux function.}\label{fig:cont}
\end{figure}

\begin{figure}
\centering
\includegraphics[width=\textwidth]{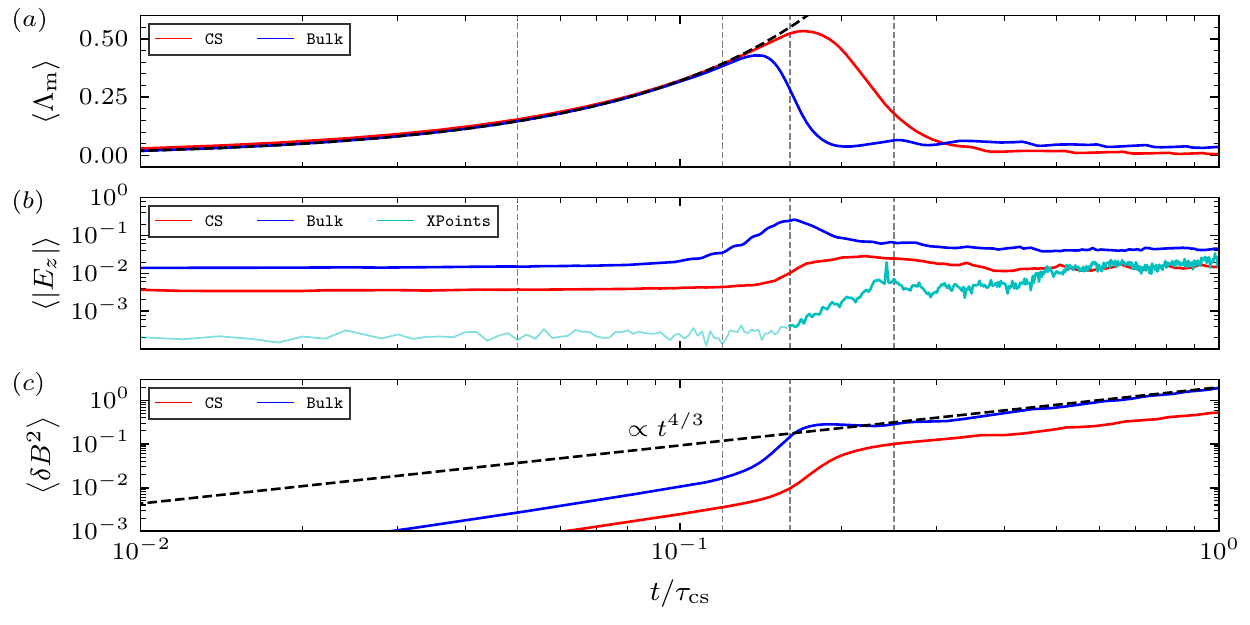}
\caption{($a$) Mirror instability threshold $\langle\Lambda_{\rm{m}}\rangle$, ($b$) out-of-plane electric field $\langle |E_z| \rangle$, ($c$) and magnetic-field fluctuation energy $\langle\delta B^2\rangle$ as a function of time from the fiducial simulation averaged over different parts of the simulation. Gray vertical lines indicate the times plotted in \cref{fig:cont}. The black dashed line in panel ($a$) represents double-adiabatic growth (see equation~\eqref{eqn:Deltap}), while the black dashed line in panel ($c$) shows the predicted secular growth ${\propto}t^{4/3}$ of the mirror fluctuations outside of the CS.
}\label{fig:fidu}
\end{figure}

\begin{figure}
\centering
\includegraphics[width=\textwidth]{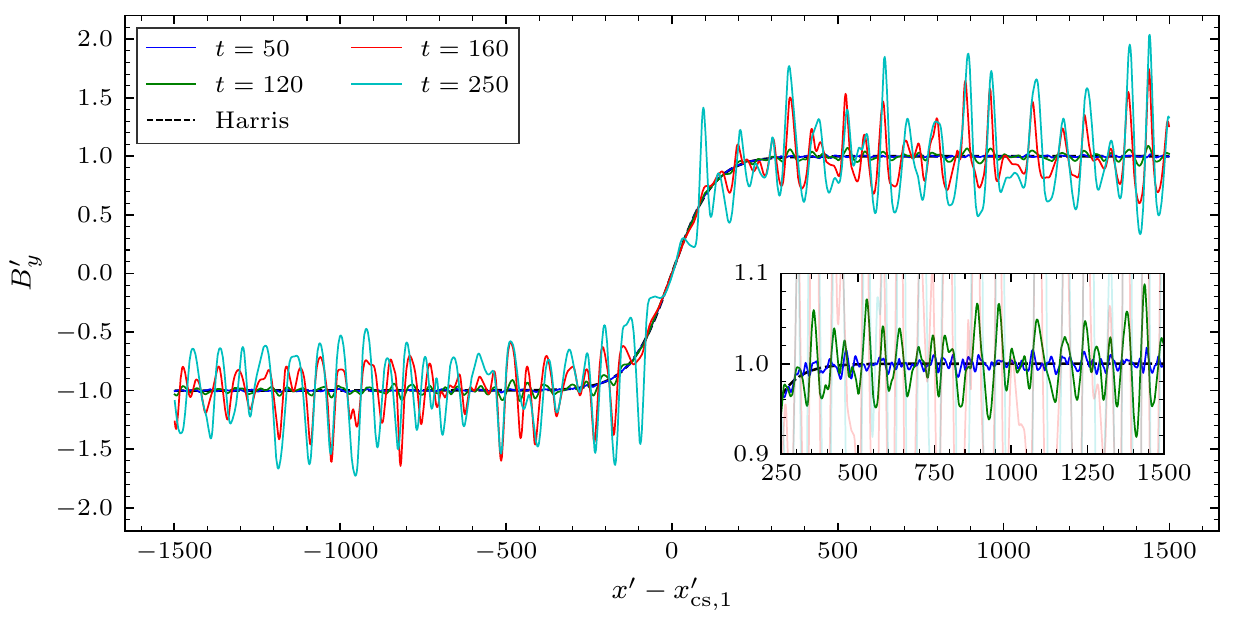}
\caption{Reconnecting field in the co-moving coordinates $B_y^\prime$ as a function of $x^\prime$ centred on the CS at $x=x_{\rm cs,1}$ at the four different times shown in \cref{fig:cont}. The black dashed line shows the profile of an undisturbed Harris sheet; the inset plot provides a zoom-in at the first two times plotted. Small-scale disturbances of the Harris profile caused by the mirror instability are evident.}\label{fig:bxfl}
\end{figure}

\begin{figure}
\centering
\includegraphics[width=\textwidth]{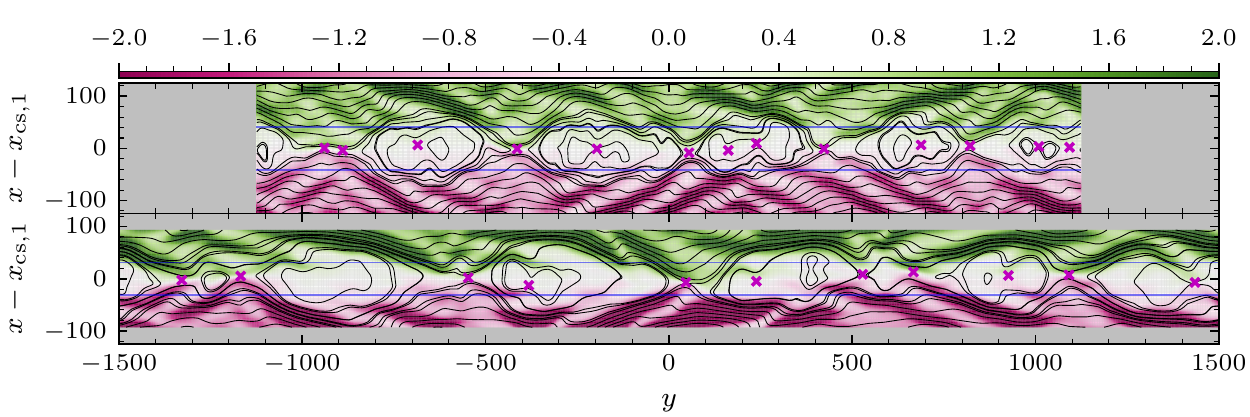}
\caption{Time slices centred about the CS at $x=x_{\rm cs,1}$ in our fiducial simulation at $t=500$ and $1000$, similar to \cref{fig:cont} but with the quantity $B_y$ plotted instead as the colour contours. These panels exhibit X-point collapse into a Y-point geometry and, in the lower panel, plasmoid formation.}\label{fig:finalt}
\end{figure}

We describe the evolution in our fiducial run using a set of four figures. Figure \ref{fig:cont} shows slices centred about a CS in our fiducial simulation taken at four different times in the evolution. All quantities, including fluctuations in the reconnecting field $\delta B_y$ (colour) and isocontours of the flux function (black lines), have been translated into the stationary lab frame; note the geometric thinning and lengthening of the CS in time. \Cref{fig:fidu} provides the time evolution in the lab frame of the mirror instability parameter $\langle\Lambda_{\rm m}\rangle$, the out-of-plane component of the electric field $\langle |E_z| \rangle$, and the energy of the magnetic fluctuations $\delta B^2(t) \doteq \langle|\bb{B}(t,x,y)-\bb{B}_{\rm r}(t,x)|^2\rangle$, each of which has been averaged separately over the \texttt{CS} and the \texttt{Bulk}. In the second panel displaying $\langle |E_z| \rangle$, we also present this value averaged over all X-point locations identified by our watershed segmentation algorithm, labelled by \texttt{XPoints}. The early values of this average are rendered transparent, since the inferred X-points during this period are strongly influenced by the low signal-to-PIC-noise ratio. Cuts of the reconnecting field in the code frame, $B'_y$, about the null line at $x'=x'_{\rm cs,1}$ are shown in \cref{fig:bxfl} for the same four times as in \cref{fig:cont}. For this figure, the code frame is chosen so that the Harris profile remains stationary and the growth of mirror fluctuations can be more easily seen. Finally, \cref{fig:finalt} supplements \cref{fig:cont} by providing the lab-frame reconnecting field $B_y$ (colour) and flux-function contours (black lines) at later times, $t=500$ and $1000$, by which point the X-points have begun to collapse into thin secondary CSs.

We begin with \cref{fig:cont}($a$), corresponding to $t = 50$. At this point, the CS has already thinned by a factor of $1.05$, causing a $5$ per cent increase in the strength of the reconnecting magnetic field. Due to the approximate conservation of adiabatic invariants, pressure anisotropy builds up in the system (\cref{fig:fidu}($a$)). While formally mirror-unstable, the accumulated anisotropy is too small at this time for the growth of the mirror instability to outpace the thinning of the CS, and the pressure anisotropy continues to increase positively following the predicted double-adiabatic evolution (see equations~\eqref{eqn:Deltap} and \eqref{eqn:Lambdam}),
\begin{equation}\label{eqn:Lambdam_pred}
    \Lambda_{\rm m}(t) = [\rmGamma(t)]^3 - 1 - \frac{\rmGamma(t)}{\beta_{i0}} \qquad \textrm{(double-adiabatic)},
\end{equation}
indicated by the dashed line in \cref{fig:fidu}($a$).

\Cref{fig:cont}($b$) corresponds to $t = 120$, by which time structures in the reconnecting field, reminiscent of the mirror instability, can be seen outside of the CS where the magnetic field is strongest. Subtle wrinkling of the field lines can also be seen in the profile of $B_y^\prime$ shown in \cref{fig:bxfl}. It is around this time that the pressure anisotropy in the \texttt{Bulk} region begins to depart from the double-adiabatic prediction and the magnetic-field fluctuations start their exponential growth in \cref{fig:fidu}($c$). In contrast, $\Lambda_{\rm m}$ is still rising steadily within the contracting CS, due to the weaker magnetization there delaying the growth of mirror modes.

As the expansion proceeds (\cref{fig:cont}($c$) at $t = 160$), significant structures in $\delta B_y$ are observed both inside and outside of the CS. In the \texttt{Bulk} plasma, $\Lambda_{\rm{m}}$ is decreasing (\cref{fig:fidu}) as the growth of mirror fluctuations depletes the pressure anisotropy. These fluctuations then grow secularly with $\langle \delta B^2\rangle \propto t^{4/3}$, as expected for the nonlinear evolution of the mirror instability (see \S\ref{sec:mirror}). Simultaneously, the profile of the reconnecting field away from the CS becomes noticeably disturbed (\cref{fig:bxfl}). Within the CS, the mirror instability also starts to grow at this time, as indicated by the stalled increase in $\Lambda_{\rm m}$ (\cref{fig:fidu}). At this stage, X-point locations near the neutral line can be distinguished using the algorithm described in section~\ref{sec:watershed}, indicating that the tearing instability has also been triggered within the CS. It is particularly notable that the characteristic spacing of these X-points not only is comparable to the parallel wavelength of the mirror fluctuations just outside of the neutral line, but also corresponds to a value of $\Delta'$ that would be negative given an undisturbed Harris profile. In other words, in the absence of the mirror instability, the observed tearing modes would still be stable. We revisit this assertion quantitatively in section~\ref{sec:compvar}.

In the final panel of \cref{fig:cont} at $t = 250$, the mirror instability in the \texttt{Bulk} region is saturated at $\langle\delta B^2\rangle \sim 0.3$ and tearing inside of the CS is fully non-linear. Reconnection is well underway and the magnetic islands have grown to the expected width of the CS and begun to merge. The consequent disturbances to the Harris profile seen in \cref{fig:bxfl} indicate large fluctuations both inside and outside of the CS. Beyond this time, $\langle |E_z| \rangle$ around the X-points (i.e., the reconnecting electric field) steadily increases in time, ultimately attaining a value ${\gtrsim}10^{-2}$, similar to that found in prior hyper-resistive MHD simulations of reconnection \citep[e.g.,][]{hbf13}. In contrast, the value of $|E_z|$ averaged over the \texttt{CS} region slowly decreases as the mirrors near the edge of the \texttt{CS} region are displaced by the growing magnetic islands, as seen in \cref{fig:finalt}($b$). In \cref{fig:finalt}($a$) at $t=500$, we also see the formation of Y-points with a Sweet--Parker-like profile, one of which (near the left boundary of this figure) eventually disrupts due to plasmoid formation at $t=1000$ in panel ($b$).

\subsection{Varying the CS formation time scale $\tau_{\rm cs}$}\label{sec:compvar}

Having described the overall evolution using our fiducial run, we now vary the CS formation time $\tau_{\rm cs}\in\{125,250,500,1000,2000\}$. At fixed $a_0=125$, this is equivalent to varying the Alfv\'{e}n Mach number $M_{\rm A0}$ of the driving flow ${\in}\{1,0.5,0.25,0.125,0.0625\}$. The goals are to test the scalings of $\Lambda_{\rm m,max}$ and $t_{\rm m,reg}$ against the theory, to demonstrate that tearing modes are always triggered on a length scale that correlates with the wavelength of the mirror instability, and to verify that such tearing modes would be stable were it not for the distortions in the Harris-sheet profile caused by the mirror fluctuations.

\begin{figure}
\includegraphics[width=\linewidth]{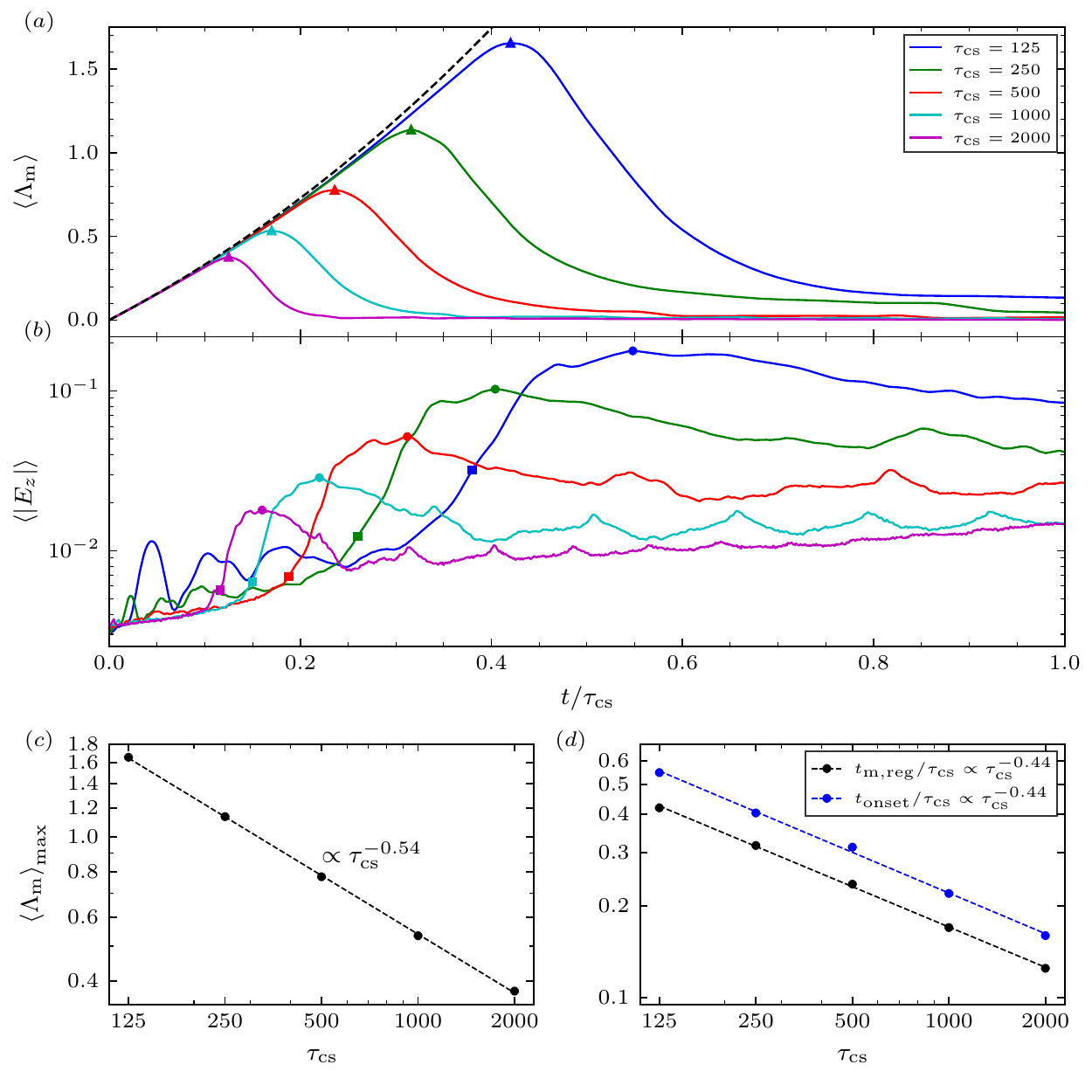}
\caption{Evolution of ($a$) $\Lambda_{\rm{m}}$ and ($b$) $E_z$ averaged over the \texttt{CS} region for different $\tau_{\rm cs}$. Maximum values of $\Lambda_{\rm m}$ are marked by triangles; the approximate starts of exponential growth in $\langle|E_z|\rangle$ for the determined X-point locations are marked by squares; and the maximum values of $\langle |E_z|\rangle$ in \texttt{CS} are marked by circles. The dashed line in panel ($a$) corresponds to \eqref{eqn:Lambdam_pred}. ($c$) Power-law scaling of $\Lambda_{\rm m,max}$ with respect to $\tau_{\rm cs}$. ($d$) Times at which the pressure anisotropy starts to be regulated by the mirror instability (black points) and at which reconnection onsets (blue points). The theoretical prediction for these dependencies in the asymptotic limit of large scale separation is $\Lambda_{\rm m,max} \propto t_{\rm m,reg}/\tau_{\rm cs}\propto \tau^{-0.5}_{\rm cs}$ (see \S\ref{sec:mirror}).
}\label{fig:mirrezcom}
\centering
\end{figure}

First, in \cref{fig:mirrezcom} we show the evolution of $\langle\Lambda_{\rm m}\rangle$ within the \texttt{CS} region for different $\tau_{\rm cs}$ (panel $a$), as well as how its maximum value, $\langle\Lambda_{\rm m}\rangle_{\rm max}$, and the `regulation' time at which its maximum value is reached, $t_{\rm m,reg}$, scale with $\tau_{\rm cs}$ (panels $c$ and $d$). As $\tau_{\rm cs}$ is increased, $\langle\Lambda_{\rm m}\rangle_{\rm max}$ decreases, consistent with the argument that the mirror instability requires less excess pressure anisotropy to outpace the CS formation when $\tau_{\rm cs}$ is large. In section~\ref{sec:mirror}, we argued that both $\Lambda_{\rm m,max}$ and $t_{\rm m,reg}$ should scale as ${\propto}\tau^{-0.5}_{\rm cs}$ in the asymptotic limit. The results in panels ($c$) and ($d$) are consistent with this scaling, the slight departures being because $t_{\rm m,reg}$ is not ${\ll}\tau_{\rm cs}$.

\begin{figure}
\includegraphics[width=\linewidth]{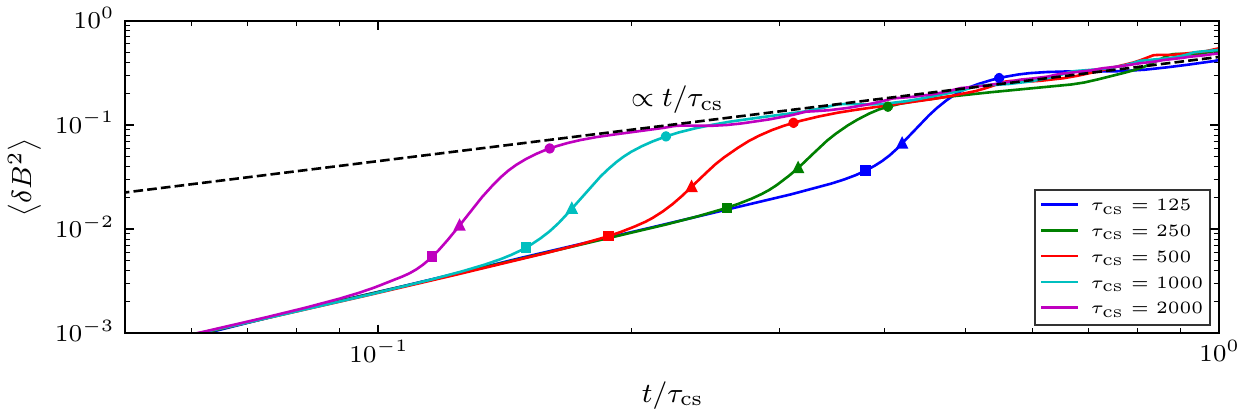}
\caption{Evolution of magnetic fluctuation $\delta B^2$ in the \texttt{CS} region for different $\tau_{\rm cs}$. The time of $\langle|E_z|\rangle$ linear growth, the maximum value of $\langle\Lambda_{\rm{m}}\rangle$, and the maximum value of $\langle|E_z|\rangle$ are marked by the square, triangle, and circle, respectively.}\label{fig:dbcom}
\centering
\end{figure}

We now turn to the evolution of $E_z$ within the CS as a function of $\tau_{\rm cs}$, shown in \cref{fig:mirrezcom}($b$). Both mirror and tearing instabilities contribute to the growth of $E_z$ in the \texttt{CS} region. By itself, the mirror instability's contribution to $E_z$ should peak at the same time as does $\Lambda_{\rm{m}}$ and then slowly decay, similar to the behaviour shown in \cref{fig:fidu}($b$) for the {\tt CS} region. Instead, we observe secondary growth of $E_z$ beyond its initial exponential growth, which we attribute to the onset of tearing. This association is strengthened by examining $E_z$ evaluated at the X-points, whose linear (exponential) growth starting at time $t_{\rm lin}$ gives way to secular growth at the same moment that $E_z$ averaged over the {\tt CS} region peaks for a second time. Given the statistical advantages of the \texttt{CS}-averaged values, we use this well-defined second peak as a proxy for the time at which reconnection `onsets'. Denoting this time as $t_{\rm onset}$, we found in all cases a temporal ordering of $t_{\rm{lin}} < t_{\rm m,reg} < t_{\rm onset}$.

This temporal ordering confirms that both instabilities grow at approximately the same time, with the tearing instability transitioning from its linear stage to its non-linear stage after the {\tt CS} pressure anisotropy begins to be regulated by the mirror instability. Note from \cref{fig:mirrezcom}($d$) that both the mirror-regulation and onset times have the same dependence on $\tau_{\rm cs}$. A comparison of the fluctuation energy shown in \cref{fig:dbcom} indicates that the growth of both instabilities rapidly increases between $t_{\rm{lin}}$ (squares) and $t_{\rm onset}$ (circles), followed by secular growth that appears to be linear in time and independent of $\tau_{\rm cs}$.

\begin{figure}
\includegraphics[width=\linewidth]{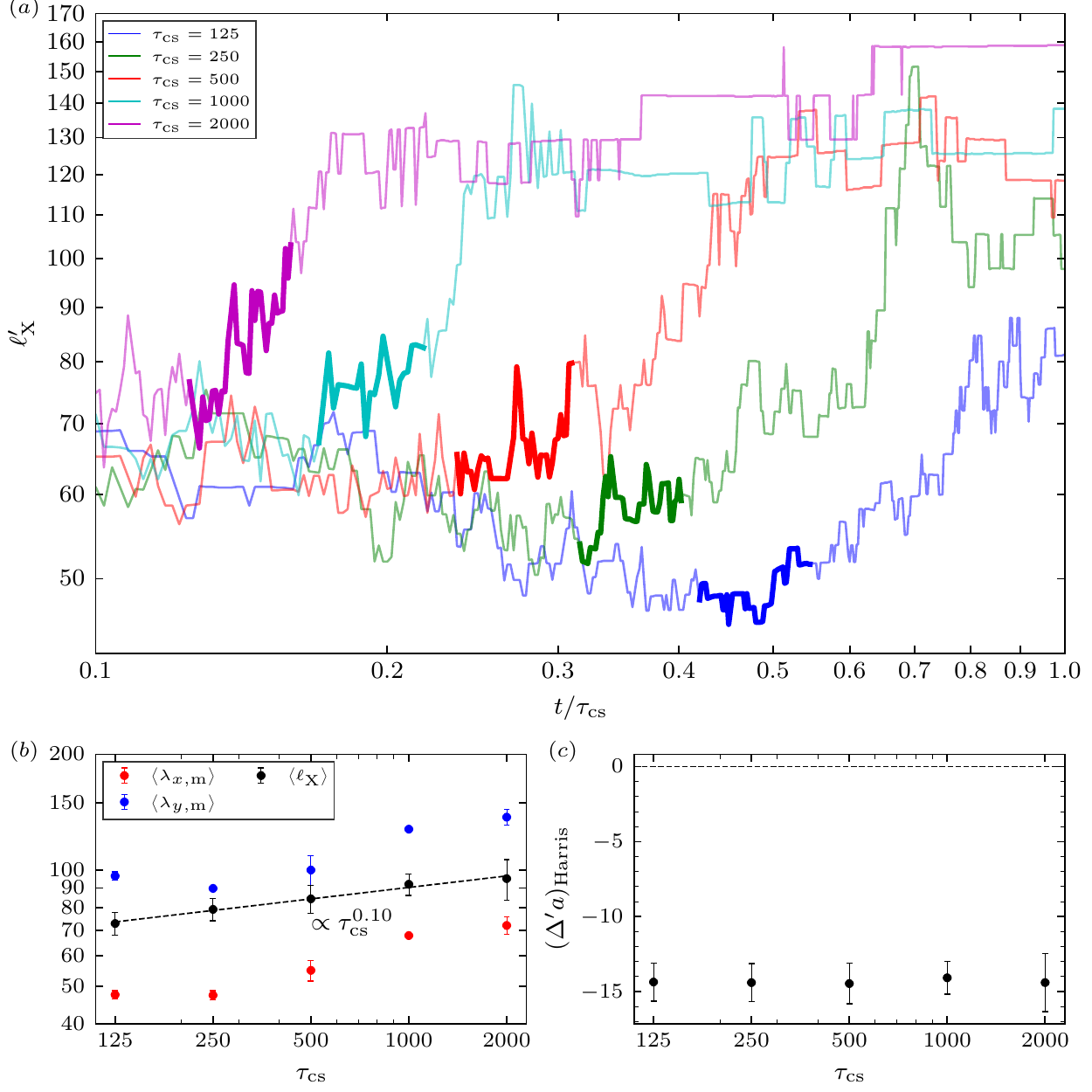}
\caption{($a$) Evolution of the average X-point separation measured in the code frame, $\ell'_\mrm{X}$, for different $\tau_{\rm cs}$, as inferred using our watershed algorithm. ($b$) Time average and standard deviation of X-point separation measured between $t_{\rm m,reg}$ and $t_{\rm onset}$ as a function of $\tau_{\rm cs}$; also shown are the dominant wavelengths of $B_y(x,y)$ measured along and perpendicular to the reconnection field during that time interval in the \texttt{Bulk} region, representing the characteristics wavelengths of mirror instability. These quantities have been transformed into the physical lab frame. ($c$) Harris-sheet $\Delta' a$ evaluated using the average $\ell_\mrm{X}$ shown in panel ($b$); note that it is negative, indicating that these tearing modes would be stable if it were not for the influence of the mirrors on the CS profile.}\label{fig:xsep}
\centering
\end{figure}

To further emphasize the impact of the mirror instability on the tearing stability of the CS, in \cref{fig:xsep} we provide an analysis of the mean separation between the X-points that form on the neutral line by tearing and its relationship to the characteristic wavelengths of the mirror instability. Panel ($a$) displays this mean separation as measured in the code frame and identified by our watershed algorithm, $\ell'_\mrm{X}$, as a function of time for different $\tau_{\rm cs}$. Here, the code frame is used because the Lagrangian advection of the islands by the driving flow makes $L/\ell_\mrm{X}$ approximately a constant during the linear growth phase of tearing; in this situation, $\ell'_\mrm{X}$ is proportional to the inverse tearing-mode number, $N^{-1}$. The thickened portions of the lines correspond to the time period between the peaking of $\Lambda_{\rm{m}}$ and the second peak of $E_z$ in the \texttt{CS} region, with the first moment corresponding to the onset of the mirror instability and the latter coinciding with the transition of the tearing instability from its linear phase to its non-linear phase. It is during this time interval that we compute an average value for $\ell_\mrm{X}$, which we associate with the characteristic tearing-mode wavelength and plot in panel ($b$) against $\tau_{\rm cs}$. We consider the value of $\ell'_\mrm{X}$ measured prior to the highlighted interval to be unreliable, because the field perturbation amplitudes caused by the tearing modes are not yet sufficiently large relative to the PIC noise to make the watershed algorithm accurate (i.e., there are many spuriously identified X-points). The rapid fluctuations seen in the curves are due to X-points that are transiently identified by the algorithm; at later times, they correspond to X-points that occur between two magnetic islands that are merging.

Two important things are noticeable in panel ($a$). First, larger values of $\tau_{\rm cs}$ result in a larger mean separation between X-points (i.e., smaller $N$). To place these values in context, note that a mode number of $N=100$ corresponds in these simulations to $\ell'_{\rm X} \simeq 94$, a value comparable to that measured during the linear phase of tearing in the simulation with $\tau_{\rm cs} = 2000$. Secondly, the X-point separations increase rapidly after the linear stage (corresponding to islands merging) and ultimately level off (indicating that mergers have slowed dramatically and the islands are mostly advected with the flow).

In panel ($b$) we compare the average X-point spacing during the linear phase of tearing, $\langle \ell_{\rm X} \rangle$, with the characteristic $x$ (i.e., perpendicular to $\bb{B}_{\rm r}$) and $y$ (i.e., parallel to $\bb{B}_{\rm r}$) wavelengths of the mirror fluctuations as measured in the {\tt Bulk} region. All three quantities are presented in the lab frame. To compute the characteristic mirror wavelength, we find the mode number at which the {\tt Bulk} energy spectrum $|B_y(k_x,k_y)|^2$ is maximal and average the implied $x$ and $y$ wavelengths using the same time interval over which $\ell_{\rm X}$ is averaged. (Recall that the \texttt{Bulk} region only contains mirror fluctuations.) The resulting characteristic mirror wavelengths are denoted by $\langle\lambda_{y,{\rm m}}\rangle$ and $\langle\lambda_{x,{\rm m}}\rangle$; their respective values are plotted as the red and blue points, with the error bars indicating their standard deviations associated with the time averaging. Note that the mirror modes go to longer wavelengths as $\tau_{\rm cs}$ increases, similarly to $\langle\ell_{\rm X}\rangle$. Moreover, $\langle \lambda_{x,{\rm m}}\rangle < \langle\ell_{\rm X}\rangle < \langle\lambda_{y,{\rm m}}\rangle$, consistent with the theoretical arguments made in section~\ref{sec:mirrorinfest}. 

Finally, panel ($c$) shows the value of the Harris-sheet $\Delta' a$, denoted by $(\Delta' a)_{\rm Harris}$, evaluated using $k_{\rm t}=2\pi/\ell_\mrm{X}$ and averaged over the same time interval used to calculate $\langle\ell_\mrm{X}\rangle$. For all values of $\tau_{\rm cs}$, we find that $(\Delta'a)_{\rm Harris}<0$, i.e., the observed tearing modes would be stable if not for the excitation of the mirror instability and consequent wrinkling of the CS on small scales. This is perhaps the clearest quantitative evidence for mirror-stimulated tearing.

\subsection{Wide Sheet} \label{sec:widesheet}

\begin{figure}
\includegraphics[width=\linewidth]{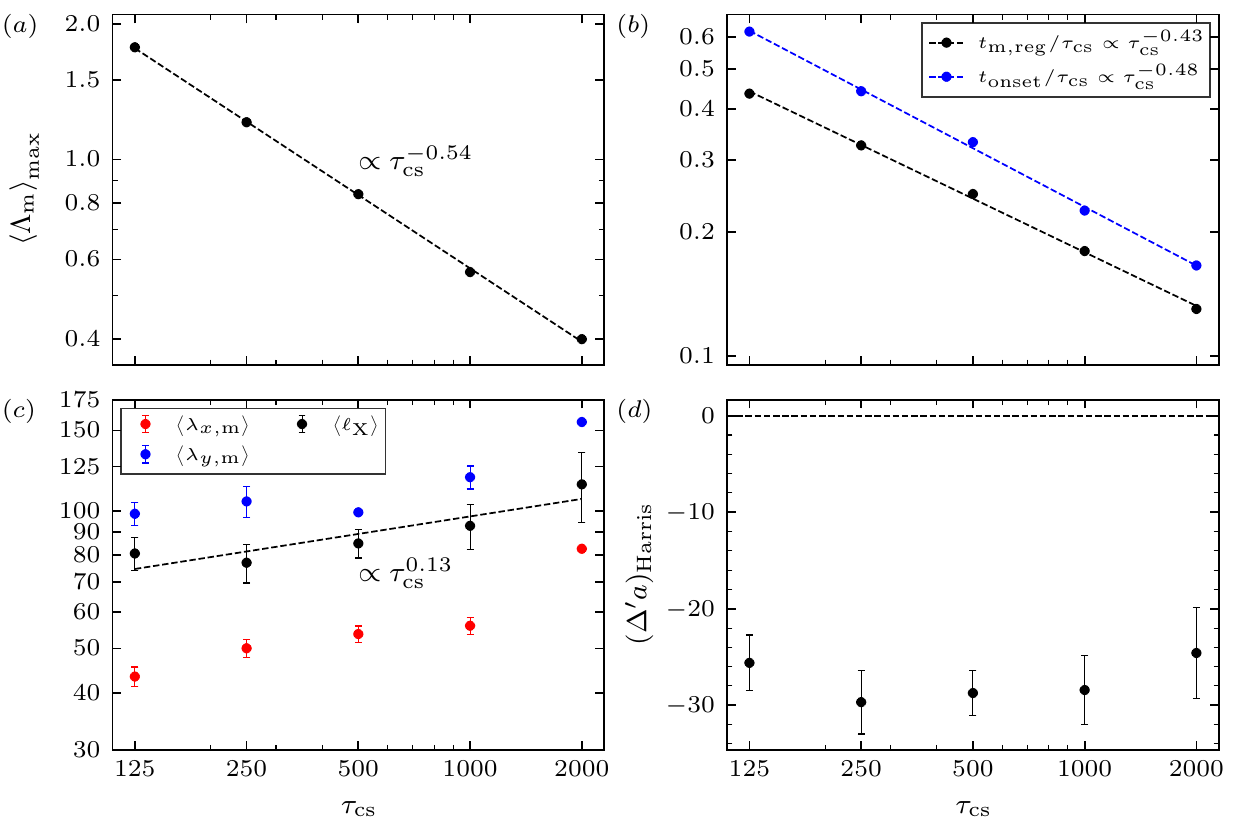}
\caption{As in figures \ref{fig:mirrezcom} and \ref{fig:xsep}, but for a CS with $a_0=250$.}\label{fig:wcom}
\centering
\end{figure}

We additionally ran simulations in which the initial CS width $a_0 = 250$, a value double that used in our fiducial run. The CS formation time is varied as in \cref{sec:compvar}, with $\tau_{\rm cs}\in\{125,250,500,1000,2000\}$. As a result, the initial Alfv\'{e}n Mach number $M_{\rm A0}$ (see \eqref{eqn:Mach}) and magnetization $a_0/\rho_{i0}$ are larger, while $\Delta'<0$ initially for all modes. In order to maintain the same level of interaction between the two CSs, we increased the system size to have $L_x = 12000$ and $N_x = 5376$ while keeping all other parameters constant.

The main results from these runs are shown in \cref{fig:wcom}, including the maximum value of $\Lambda_{\rm m}$ in the \texttt{CS} region (panel $a$), the mirror-regulation and tearing-onset times (panel $b$), the mean X-point separation compared to the characteristic mirror wavelengths (panel $c$), and the Harris-sheet $\Delta' a$ evaluated at $k_{\rm t}=2\pi/\ell_\mrm{X}$ (panel $d$), all as functions of $\tau_{\rm cs}$. The scalings shown in panels ($a$) and ($b$) are nearly identical to those seen in \cref{fig:mirrezcom}($c$) and ($d$), and are consistent with the expected scalings for the mirror-regulation and tearing-onset times. Likewise, panel ($c$) exhibits a $\tau_{\rm cs}$ scaling for the mean X-point separation similar to that found in \cref{fig:xsep}($b$), with $\langle \lambda_{x,{\rm m}}\rangle < \langle\ell_{\rm X}\rangle < \langle\lambda_{y,{\rm m}}\rangle$. Once again, these values of $\ell_\mrm{X}$ imply a tearing-mode wavenumber that would be stable in an undisturbed Harris sheet; in fact, the values of $(\Delta' a)_{\rm Harris}$ shown in panel ($d$) are nearly twice as negative as those found when $a_0=125$. Since the onset times are approximately equal to those found in \cref{sec:compvar}, this confirms that the mirror instability is the main instigator of tearing and the subsequent onset of reconnection.

\section{Discussion}\label{sec:summary}

We have used hybrid-kinetic simulations to show that a steadily thinning CS in a collisionless, magnetized plasma accumulates pressure anisotropy in the inflowing fluid elements, and that this anisotropy quickly goes mirror-unstable at sufficiently large $\beta$. The subsequent ion-Larmor-scale wrinkling of the CS modifies the profile of the reconnecting field in a way that dramatically reduces its effective thickness and, thereby, its stability to tearing modes (quantified through $\Delta'$). Simultaneously, the rapid growth of the mirror fluctuations directly stimulates tearing modes by providing a nonlinear seed. As a result, tearing onsets earlier and on smaller scales than it would have without the mirrors, thereby placing a tighter upper limit on the aspect ratio of any forming CS. By varying the CS formation time $\tau_{\rm cs}$, we find that the reconnection onset time $\tau_{\rm onset}$ is approximately proportional to $\tau^{1/2}_{\rm cs}$, as the mirror instability grows proportionally earlier in simulations with slower compression. In other words, the ratio $\tau_{\rm onset}/\tau_{\rm cs}$ decreases with increasing scale separation. Increasing the initial CS thickness $a_0$ returns hardly any change in the outcome: the onset times are close to those obtained at smaller $a_0$, and the unstable tearing modes are again intermediate in scale between the parallel and perpendicular wavelengths of the mirrors. In all cases, the mirror-stimulated tearing modes ultimately grow to produce multiple islands whose widths are comparable to the CS thickness.

Our work lends credence to the conclusion made by \citet{ak19} that `numerical simulations of collisionless reconnection in high-$\beta$ plasmas should not initialize with a Maxwellian plasma embedded in an equilibrium CS. Instead, the CS should be allowed to evolve, and the particle distribution function self-consistently with it.' It also provides a natural explanation for results from a recent laser-plasma experiment of driven reconnection with collisionless ions, which indicated an earlier onset of tearing having larger growth rates and significantly smaller scales than anticipated \citep{fox21}. These successes borne in mind, it is useful to place our work in the broader context of the turbulent plasma dynamo, in which chaotic large-scale fluid motions organize a growing magnetic field into a highly intermittent patchwork of long, thin, reversing structures. These `magnetic folds' may be viewed locally as CSs that, depending on their aspect ratio and the material properties of the plasma, may be susceptible to disruption by tearing \citep{galishnikova22}. In a collisionless or weakly collisional plasma, the generation of these folds involves the production of positive pressure anisotropy in the regions of low magnetic curvature \citep[e.g.,][]{rincon15,stonge18,stonge20}, in a manner qualitatively similar to the CS-formation model employed here. Assuming our results carry over to that more complicated system, the implication is that such folds will experience mirror-stimulated tearing and breakup into plasmoid-like flux ropes before they are able to thin to resistive (or electron-kinetic) scales. How electron pressure anisotropy interferes with or assists this process (e.g., by triggering electron-Larmor-scale instabilities that may result in an anomalous electrical resistivity) awaits a more general treatment of CS formation and reconnection physics than we have employed here. Fully kinetic simulations would be most informative.

%
%
\section*{Acknowledgments}
It is a pleasure to thank Andy Alt, Lev Arzamasskiy, Archie Bott, Vladimir Zhdankin, and especially Nuno Loureiro for useful conversations and the expert referee for a constructive report. This work was supported by U.S.~DOE contract DE-AC02-09CH11466 and NSF CAREER award No.~1944972, and is part of the Frontera computing project at the Texas Advanced Computing Center; it also made extensive use of the {\em Perseus} and {\em Stellar} clusters at the PICSciE-OIT TIGRESS High Performance Computing Center and Visualization Laboratory at Princeton University.

\section*{Declaration of Interests}
The authors report no conflict of interest.


\end{document}